\documentclass{article}      

\title{An Example Document}  
\author{Leslie Lamport}      
\date{January 21, 1994}      

\begin{document}

\title{A HIC Primitive Spinodal Decomposition Signature}

\author{   A.  Barra\~n\'on  
\footnote{ Universidad Aut\'onoma Metropolitana. Unidad Azcapotzalco.
Av. San Pablo 124, Col. Reynosa-Tamaulipas, Mexico City. email: bca@correo.azc.uam.mx } ;
 J. A.  L\'opez
\footnote{Dept. of Physics, The University of Texas at El Paso. El Paso, TX, 79968  }   } 

\def\rightmark{HIC Primitive Spinodal Decomposition}
\def\leftmark{A. Barra\~n\'on et al.}

\date{May, $20^{th}$,  2004}

\maketitle

\abstract

Evidence of a primitive spinodal decomposition has been obtained for central Ni+Ni
Heavy Ion Collision, since  higher order charge correlations show a peak when four
 fragments of size equal to 6 are produced with an excitation of 4.75 MeV. This can
 be considered as a signature of a primitive breakup in equal sized fragments with
 a privileged fragment size. This computational result confirms other experimental
 and theoretical evidences about spinodal decompostion in HIC.

\section{ Introduction}

Heavy Ion Collisions are expected to experience a liquid-gas phase transition due to the
 specifics of the internucleonic interaction, which is attractive in both the long and
 intermediate ranges and repulsive in the short range. It is possible that 
a wide zone of phase space is explored when two nuclei collide, including a region
 where liquid and gas phases coexist, namely the spinodal region. At this spinodal
 region, incompressibility is negative and uniform nuclear matter is unstable, leading
 to multifragmentation due to the increase of  density fluctuations \cite{1} \cite{2}. 
Dynamical simulations based on Boltzmann equation, such as Landau-Vlasov (LV), 
Boltzmann-Uehling-Uhlenbeck (BUU) or Boltzmann-Nordheim-Vlasov (BNV), 
describe the time evolution of the density of a one-body system, ignoring those
 correlations whose order is larger than the order of binary correlations, neglecting
 fluctuations around the mean trajectory of the system, which altogether comes out to 
be quite inconvenient to study the spinodal instability zone \cite{1}. As shown by
 Guarnera et. al. \cite{3}, a spinodal decomposition produces a "primitive breakup" 
where equal sized fragments have a privileged fragment size, which is related to the
 wave lenghts of the most unstables modes of nuclear matter. Tabacaru et. al. have 
obtained reduced velocity correlations between fragments and did not find a bubble-like
 profile, which excludes surface instabilities that might cause multifragmentation. 
They also computed higher order charge correlations, defined as the ratio between
 the number of correlated fragments and the number of uncorrelated fragments, 
obtaining evidence of a privileged production of equal sized fragments \cite{4}. 
Chomaz et. al.  have introduced a scenario inspired on experimental data,  where
 a gently compressed systems expands and reaches thermal equilibrium approximately
 at the time when the system enters into the spinodal region. At this moment,
 density fluctuations break up the system  into several hot fragments and particles.
 When fragments are released from the nuclear force, configuration frozens and 
fragments only interact with each other via coulomb force. At this moment, system
 has explored so much phase space that it can be described by statistical models and
 there is no contradiction between statistical or dynamical approximations. Statistical
 approximations describe the time evolution of the system and the phase
 diagram. Meanwhile, dynamical approximations start up in phase diagram and are 
more related to the thermodynamics of non extensive systems. Barra\~n\'on et. al. have  
obtained  computational evidence about the inverse relation between entropy and the
residual size, using LATINO dynamical model to study the spinodal decomposition 
region of central Ni+Ni HIC at intermediate energies  \cite{5} . In this very study,
 evidence is obtained about a primitive spinodal decomposition for Ni+Ni central HIC at 
intermediate energies. 

\section{ Methodology.}

Heavy Ion Collisions were simulated using LATINO semiclassical model where
 binary interaction \cite{6} is reproduced with a Pandharipande potential built up
  of Yukawa potentials linear combinations, whose coefficients are designed to
 both reproduce nuclear matter ground state properties and to fulfill Pauli exclusion
 principle \cite{7}:
\begin{equation}
V_{nn}=V_{pp}=V_{0} \bigl( \frac {e^{- \mu_0 r} } {r} -  \frac {e^{- \mu_0 r_C} } {r_c} \bigr)
\end{equation}
and:
\begin{equation}
V_{np}= V_{r} \bigl( \frac {e^{- \mu_r r} } {r} -  \frac {e^{- \mu_r r_C} } {r_c} \bigr) - V_{a} \bigl( \frac {e^{- \mu_a r} } {r} -  \frac {e^{- \mu_a r_a} } {r_a} \bigr)
\end{equation}

 Clusters are identified with an Early Cluster Recognition Algorithm
 that optimizes configurations in energy space. A most bound partition is obtained
 minimizing the sum of energies of the clusters belonging to each partition:

\begin{equation}
\{ C_i \} = arg min \bigl[ E_{ \{ C_i \} } = \sum_i E_{int}^{C_i} \bigr]
\end{equation}
where the energy of each cluser is given by : 

\begin{equation}
 E_{int}^{C_i} =\sum_i \bigl[ \sum_{ij \in C_i} K_{j}^{CM} + \sum_{j,k \in C_i, j \le k} V_{jk} \bigr]
\end{equation}
where the first sum includes the partition clusters, $K_{j}^{CM}$ is the kinetic energy of the
 particle j measured in the center of mass of the cluster containing particle j, 
and $V_{ij}$ is the internucleonic potential. The algorithm uses "simmulated annealing" 
to find the most bound partition and optimizes the partition in energy space.

   Projectile energy is in the range of 600 to 2000 MeV and system evolves until its
 microscopic composition is rather frozen though some monomers are ejected. This time 
can be identified using the Persistence Microscopic Coefficient, defined as the probability 
that two particles belonging to the partition X remain bound in partition Y:
\begin{equation}
 P \bigl[ X,Y \bigr]= \frac{1}{ \sum_{cluster} n_i} \sum_{cluster} \frac{  n_i a_i } { b_i }   
\end{equation}
where $b_i$ is equal to the number of pairs of particles belonging to the cluster  $C_i$ 
of partition $X$ while $a_i $ is equal to the number of particle pairs belonging to cluster 
$C_i$ of partition $X $ that also belong to a given cluster $C'_i$ of partition $Y$. 
$n_i$ is the number of particles in cluster $C_i$.

 Higher order charge correlations were introduced by Moretto et. al. \cite{8} 
and are given by: 

\begin{equation}
 \frac{ Y \bigl( \Delta Z, <Z> \bigr ) }{  Y' \bigl( \Delta Z, <Z> \bigr) } \biggr|_M
\end{equation}
where $ Y \bigl( \Delta Z, <Z> \bigr )$ is equal to the number of fragments produced 
for given values of $\Delta Z $ and $ <Z> $. 

\section{Results}

Higher order charge correlations were obtained for central HIC Ni+Ni. 
These higher order correlations show a peak for a fragment size equal 
to 6, with four equal sized fragments produced with this privileged size
 and an excitation equal to 4.75 MeV. This can be considered a
 primitive breakup signature in equal sized fragments of a privileged
 size. Hence, this dynamical study supports other experimental and theoretical
 evidences of spinodal decomposition in HIC, such as the
 experimental study reported by Borderie et. al.
\cite{9} where a fosil spinodal decomposition signature for HIC collision
 $^{129}Xe+ ^{nat}Sn$ at  32 AMeV was obtained. At that very experiment, 
liquid-gas coexistence was observed and evidence was obtained of a first
 order phase transition for a non extensive particle system. This HIC phase
 transition  has been identified in previous studies with several signatures, 
namely Campi scattered plotts \cite{10} as well as fragment size distribution
 power laws \cite{11}. 
 
\section{Conclusions.}

Once LATINO dynamical model was applied, higher order charge 
correlations provided computational evidence about the spinodal 
decomposition in the early stage of fragmentation for central HIC
 Ni+Ni at intermediate energies. This confirms previous evidence reported by 
others of a fosil spinodal decomposition, indicating liquid-gas
 coexistence and a first order phase transition for non extensive 
central HIC Ni+Ni. Authors are grateful for the hospitality of the
 Instituto de F\'{\i}sica at UNAM. A.B. acknowledge partial funding
 from UAM-A and ready acces to the computational resources of the 
Intensive Computing Lab at UAM-A. Work supported by 
National Science Foundation (PHY-96-00038).


\begin{thebibliography}{10}
\bibitem{1}
J. D. Frankland {\it et. al.} . Nucl.Phys. A{\bf689} (2001)940-964 
\bibitem{2} 
L.G. Moretto and G.J. Wozniak . Ann. Rev. of Nuclear and Particle Science {\bf43} (1993)379.
\bibitem{3}
A. Guarnera {\it et. al.} . Phys. Lett. B{\bf373} (1996)267. 
\bibitem{4} 
G. Tabacaru {\it et. al.}. ArXiv Preprint:nucl-th/0102015 (2001)41.
\bibitem{5}
A.Barra\~n\'on {\it et. al.}. Physical Review C {\bf69}, Number 1 (2004)1. 
\bibitem{6}
A. Barra\~n\'on {\it et. al.}. Rev. Mex. Phys. {\bf 45} (1999)110. 
\bibitem{7} 
R. Lenk {\it et. al.} . Phys. Rev. C, {\bf42} (1990)372 
\bibitem{8} 
L. G. Moretto {\it et. al.} . Phys. Rev. Lett. {\bf77} (1996)2634. 
\bibitem{9}
B. Borderie {\it et. al.} . Phys.Rev.Lett. {\bf86} (2001)3252-3255 
\bibitem{10}
A.Barra\~n\'on {\it et. al.}.Rev. Mex. Phys. {\bf47}-Sup.2, (2001)93. 
\bibitem{11}
A. Barra\~n\'on {\it et. al.}. Heavy Ion Phys. {\bf17}-1, (2003)59. 
\end{thebibliography}
\end{document}